\documentclass[aps,prl,preprint]{revtex4}
\usepackage{graphicx}
\begin{document}
\title{Strong Suppression of Electrical Noise in Bilayer Graphene \\Nano Devices}

\author {YU-MING LIN} \email{yming@us.ibm.com}
\author {PHAEDON AVOURIS} \email{avouris@us.ibm.com}
\affiliation{IBM T.\ J.\ Watson Research Center, Yorktown Heights,
NY 10598, USA}
\begin{abstract}
Low-frequency 1/f noise is ubiquitous, and  dominates the
signal-to-noise performance in nanodevices. Here we investigate the
noise characteristics of single-layer and bilayer graphene
nano-devices, and uncover an unexpected 1/f noise behavior for
bilayer devices. Graphene is a single layer of graphite, where
carbon atoms form a 2D honeycomb lattice. Despite the similar
composition, bilayer graphene (two graphene monolayers stacked in
the natural graphite order) is a distinct 2D system with a different
band structure and electrical
properties.\,\cite{McCann_PRBRapid2006,Ohta_science2006} In graphene
monolayers, the 1/f noise is found to follow Hooge's empirical
relation with a noise parameter comparable to that of bulk
semiconductors. However, this 1/f noise is strongly suppressed in
bilayer graphene devices, and exhibits an unusual dependence on the
carrier density, different from most other materials. The unexpected
noise behavior in graphene bilayers is associated with its unique
band structure that varies with the charge distribution among the
two layers, resulting in an effective screening of potential
fluctuations due to external impurity charges. The findings here
point to exciting opportunities for graphene bilayers in low-noise
applications.
\end{abstract}

\maketitle Ultra-thin graphite films have attracted strong
scientific and technological interest as truly 2D transport
systems\,\cite{Zhang_Kim_Nature2005,rise_graphene,carbon_electronics_review}
with exceptional carrier
mobilities.\cite{novoselov_nature2005,Berger_science2006} While a
single-layer 2D graphene is a zero-gap semiconductor, which is not
suitable for certain applications, a wealth of different band
structures emerge in nanoscale graphene that exhibit band gaps and
distinct electrical properties\,\cite{PhysRevB.54.17954} desirable
for various device applications, such as metallic
interconnects,\,\cite{Berger_science2006} field-effect
transistors\,\cite{zhihong_graphene2007,Han_PRL2007} and
single-charge devices.\,\cite{rise_graphene} In addition, both
experimental\,\cite{zhihong_graphene2007,impurity_scattering} and
theoretical\,\cite{ando_graphene_screening} studies have shown that
the transport properties of graphene systems are highly sensitive to
external perturbations such as adsorbed molecules on the surface or
nearby impurity charges, indicating that graphene is also
advantageous for sensor applications.\cite{gas_detection_graphene}
However, this high sensitivity of graphene devices also implies that
any uncontrolled and random perturbations in the environment would
lead to significant device current fluctuations, and contribute to
low-frequency 1/f noise. We note that while 1/f noise is ubiquitous
in solid-state electronic devices, this type of noise is of
particular significance in determining and/or limiting the
performance for nanoscale devices because its amplitude always
increases with diminishing device dimensions. Moreover, for
high-frequency applications where graphene holds great potential
because of its high mobility, this low-frequency noise can be
up-converted to induce phase noise in RF designs\,\cite{phase_noise}
and affect device performance. Despite the broad interest and
intense experimental focus on graphene, a comprehensive study on the
noise characteristics of graphene devices is still lacking.

Here we report on the fabrication and investigation of the
electrical noise characteristics of two-terminal single-layer
graphene (SLG) and bilayer graphene (BLG) nanodevices. Structurally,
carbon nanotubes are considered as a special case of rolled-up
graphene nano-ribbons, and therefore, much of the knowledge acquired
from the extensive studies on carbon nanotubes may be readily
applied here to shed light on the understanding of experimental
results. The behavior of 1/f noise in single-walled carbon nanotube
devices has been systematically studied previously, and it is found
that the origin of 1/f noise in these nano devices is dominated by
fluctuations of trap charges in the
oxide.\,\cite{Lin_Noise_Nanoletter2006,Ishigami_APL2006} The noise
level in carbon nanotube devices can be lowered by improving the
oxide quality through passivation\,\cite{ALD_nosie_APL_Shim2007}
and/or thermal annealing,\,\cite{Lin_noise_PhysE} or by removing the
oxide entirely.\,\cite{Lin_noise_nanotechnology2007} In these
efforts, the noise reduction is due to the elimination of trap
charges, regardless of the properties of the transport channel. In
this report, we demonstrate a different approach to reduce 1/f noise
in graphene devices, as illustrated by the strong suppression of 1/f
noise in graphene bilayers compared to single-layer graphene and
nanotube devices, all in the presence of similar oxide quality. This
unexpected noise behavior in graphene bilayers is associated with
its unique band structure that varies with the charge distribution
among the two layers, resulting in an effective screening of
potential fluctuations associated with external impurity charges.
These results not only point to exciting opportunities for graphene
bilayers in low-noise applications, but also provide valuable
insight for further noise reduction in nano devices.

Fig.\,\ref{device_setup}A shows a scanning electron microscopy (SEM)
image of the graphite flake used in this study, where regions with
different numbers of graphene layers can be readily identified from
the image brightness. Fig.\,\ref{device_setup}B shows the trace of
the AFM height profile measured along the arrow in
Fig.\,\ref{device_setup}A, corresponding to regions with mono-, bi-
and tri-layer graphene.  The inset of Fig.\,\ref{device_setup}C
shows an SEM image of a graphene nanoribbon device with a channel
width of 30 nm, following fabrication methods reported
previously.\,\cite{zhihong_graphene2007,Han_PRL2007}

Using the underlying Si substrate as the gate electrode,
Fig.\,\ref{device_setup}C shows the device resistance as a function
of gate voltage for two SLG and BLG devices with identical channel
geometry (see inset). We note that both devices exhibit a Dirac
point corresponding to the resistance maximum near $V_g$=0,
indicating insignificant doping effects in the fabrication process.
Compared to SLG nanodevices, the BLG device of the same geometry
always possesses a lower channel resistance and a weaker gate
dependence. For example, in Fig.\,\ref{device_setup}C, BLG and SLG
devices exhibit comparable resistance values at either sufficiently
large or negative gate voltages ($V_g\sim\pm 20$\,V), whereas the
resistance of the SLG device is more than 3 times higher than that
of the BLG at the Dirac point. The fact that the resistance ratio of
the SLG and BLG devices is not a constant as a function of $V_g$
clearly demonstrates that carrier transport in BLG devices cannot
simply be regarded as two non-interacting graphene layers placed in
parallel. Instead, the stacking order in bilayer graphene leads to a
different 2D dispersion relation from that of SLG, which should be
used to describe the transport properties in BLG.  We also note that
in Fig.\,\ref{device_setup}C, the resistance maximum of the BLG
device is less than half of their SLG counterpart, consistent with
theoretical calculations that predict a resistance ratio as large as
6 at the Dirac point for the two cases in the weak-disorder
limit.\,\cite{Koshino_bilayer_PRB2006,Cserti_minimal_cond_PRL2007}


In addition to the different gate dependence, a closer examination
of Fig.\,\ref{device_setup}C reveals that the  measured $I$-$V_g$
curve of the BLG device is much smoother than that of the SLG
device, an intriguing observation for two nano-devices having the
same dimensions and comparable resistance. In order to
quantitatively study and compare these current fluctuations, we use
a spectrum analyzer to measure the current power spectra at a dc
bias, and at first we focus on a SLG device with a channel width $W$
of 30\,nm and length $L$ of 1.7\,$\mu$m. The inset of
Fig.\,\ref{sgn_noise}A shows the measured current power spectral
density $S_I$ as a function of frequency (f) of the SLG device at a
bias $V_d$=100\,mV for two different gate voltages. The power
spectral density is found to be proportional to the current square
$I^2$ in the linear $I$-$V_d$ regime and inversely proportional to
the frequency (see Fig.\,\ref{sgn_noise}A). This frequency
dependence is characteristic of the so-called 1/f noise, which is
the dominant low-frequency noise existing in essentially all
electronic materials. The current power spectra can be expressed as
\begin{equation}
S_I=A_N\cdot {I^2}\cdot {f^\beta}, \label{eq:noise}
\end{equation}
where $\beta$ is the frequency exponent with a value close to $-1$,
to within $\pm 0.1$, and $A_N$ is the 1/f noise amplitude. We note
that this $S_I \propto I^2$ dependence in the linear $I$-$V_d$ curve
shown in Fig.\,\ref{sgn_noise}A indicates that the 1/f noise in SLG
is due to resistance fluctuations, i.e.
$S_R/R^2=S_I/I^2$.\,\cite{Lin_Noise_Nanoletter2006}

The noise amplitude $A_N$ of the SLG device is found to be dependent
on the gate voltage, as illustrated by Fig.\,\ref{sgn_noise}A, which
shows a weaker noise spectrum, $S_I/I^2$, for a gate voltage
corresponding to a lower-resistance state. Fig.\,\ref{sgn_noise}B
plots the device resistance, $R$, and the noise amplitude, $A_N$, as
a function of $V_g$, showing a positive correlation between these
two quantities. In order to understand this 1/f noise behavior, we
compare the SLG device with a single-wall carbon nanotube (CNT)
device; in terms of transport, both can be visualized as a ribbon of
single-layer graphene with a finite and narrow width. The noise
characteristics of individual nanotube devices have been studied
previously,\,\cite{Lin_Noise_Nanoletter2006,Ishigami_APL2006} and
their noise power spectrum is also described by
Eq.\,(\ref{eq:noise}). In semiconducting nanotubes, the noise
amplitude $A_N$ exhibits a strong gate dependence by more than two
orders of magnitude, as in our SLG devices, and $A_N$ is found to
follow the empirical relation $A_N=\alpha_H/N$, called
 Hooge's relation,\,\cite{hooge_PL1969} where $N$ is the total number of transport
carriers in the device channel and $\alpha_H$ is defined as Hooge's
noise parameter. It is important to point out that this noise
parameter $\alpha_H$, while assumed constant for a given device or
material, is not an absolute constant and, instead, is often used as
a technological measure for electronic device and/or materials
quality. For nanotube devices fabricated on a SiO$_2$ surface,
$\alpha_H$ is roughly $\sim 10^{-3}$, and its exact value is
determined by the gate oxide
quality.\,\cite{Lin_noise_nanotechnology2007} We note that since the
resistance of SLG devices is inversely proportional to the carrier
density, the fact that the measured resistance $R$ and noise
amplitude $A_N$ curves possess nearly identical gate dependence (see
Fig.\,\ref{sgn_noise}B), provides strong evidence that the 1/f noise
behavior follows Hooge's relation in single-layer graphene samples,
as in the case of CNTs.

From Hooge's relation for $A_N$ and the resistance of SLG devices
given by $R=(en\mu)^{-1}\cdot (L/W)$, we find
\begin{equation} A_N/R=\left( \frac{e\mu}{L^2}\right)\alpha_H,
\label{eq:2}
\end{equation}
where $\mu$ is the carrier mobility and $L$ is the device channel
length. To calculate the mobility, we utilize the gate-dependent
conductance along with the relation $n=\gamma\cdot V_g$ (where
$\gamma\approx 7.2\times 10^{10}$\,cm$^{-2}$/V for a 300\,nm SiO$_2$
layer\,\cite{novoselov_nature2005,Zhang_Kim_Nature2005}), and obtain
a field-effect carrier mobility $\mu$ $\sim 700$\,cm$^2$/Vs for the
30-nm-wide SLG device shown in Fig.\,\ref{sgn_noise}B. In
Fig.\,\ref{sgn_noise}C, we plot the noise amplitude $A_N$ as a
function of resistance, showing a linear dependence in agreement
with Hooge's relation. Based on the fitted slope $A_N/R= 3.8\times
10^{-12}$(1/$\Omega$) in Fig.\,\ref{sgn_noise}C, the Hooge's noise
parameter of the SLG device is found to be $\alpha_H\sim 1\times
10^{-3}$ (see Eq.\,(\ref{eq:2})).  To ensure the validity of this
noise analysis for graphene, Fig.\,\ref{sgn_noise}C also shows the
power exponent $\beta$ of the noise power spectra as a function of
device resistance, confirming the 1/f dependence for all gate
voltages.
 Although it may not be unexpected that
both single-wall CNT and single-layer graphene devices possess
similar 1/f noise characteristics described by Hooge's relation and
yield comparable noise parameters $\alpha_H$ in the range of
$10^{-3}$, these are importance findings with significant
implications. First, despite different contact configurations (ohmic
in SLG\,\cite{rise_graphene} vs. Schottky barrier in
CNT\,\cite{appenzeller_PRL2002}), these results clearly suggest that
in both systems the 1/f noise is dominated by the presence of the
underlying oxide, where trapping/detrapping processes are expected
to be the major sources for the 1/f noise. In particular, the
quantitatively similar noise behavior in the two systems may arise
from the similar configuration where the electrical transports are
{\it entirely} carried by mobile charges that are in direct contact
with the same oxide layer. There is, however, an important
difference between the SLG nanoribbon and the CNT devices associated
with the edge states at the channel boundary in the graphene
nanoribbon, which has been shown to affect its transport properties
and lead to a lower mobility in nanoribbon devices than that of
their un-patterned counterparts.\,\cite{zhihong_graphene2007}
However, since both CNT and SLG devices yield comparable $\alpha_H$
values, these uncontrolled edge states do not seem to affect the 1/f
noise in graphene nanoribbons. In addition, the symmetry of the
noise amplitude and resistance with respect to $n$ and $p$ branches
(see Fig.\,\ref{sgn_noise}B), where the resulting $A_N$-$R$ data
sets can be nicely fitted by one single slope, indicates that the
processes responsible for the 1/f noise does not depend on the type
of transport carrier.

\begin{figure} \center
\includegraphics[width=4in]{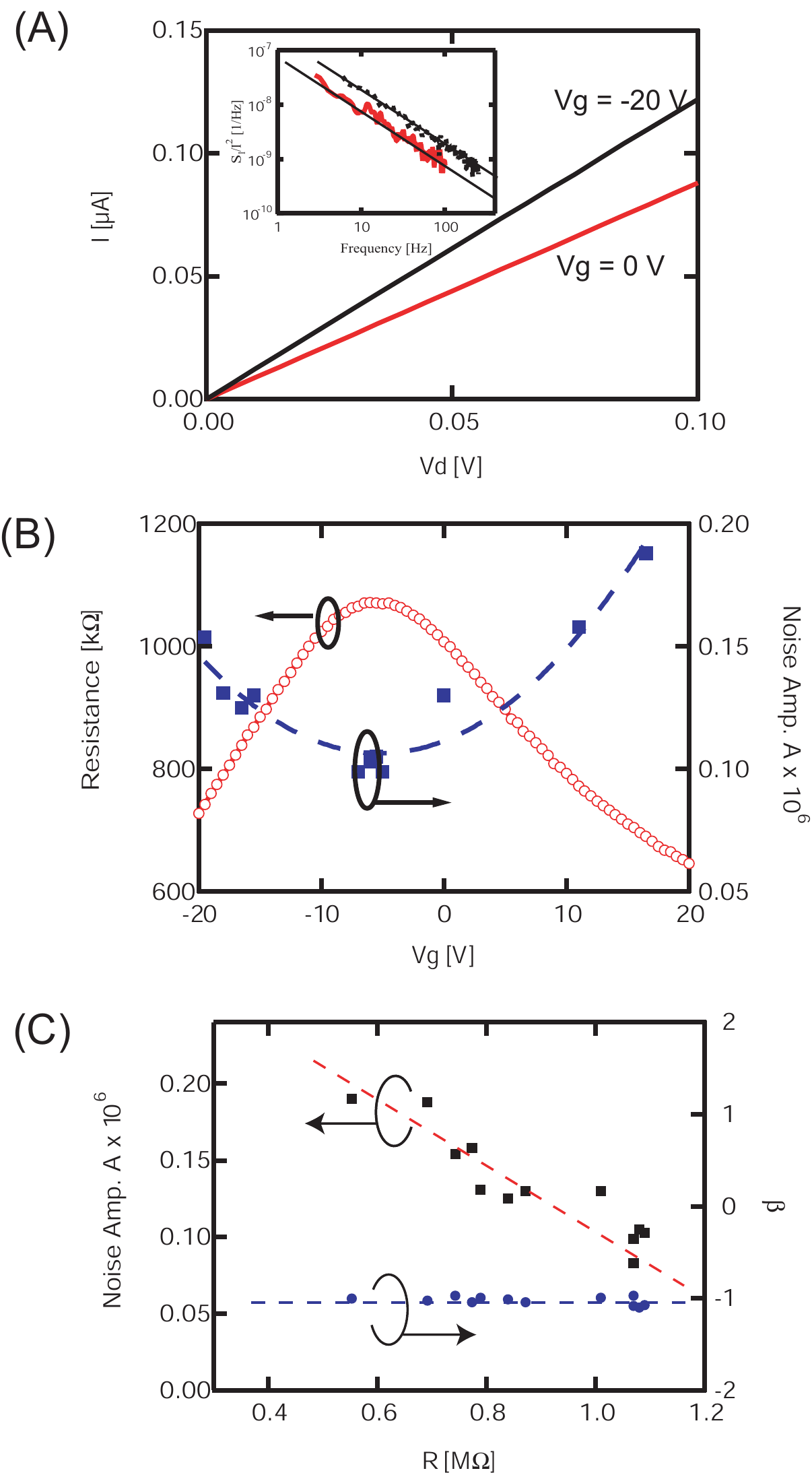}
\caption{\label{dbl_noise}}
\end{figure}

Next, we examine the noise characteristics of a bilayer graphene
nanodevice with the same channel width $W=30$\,nm and a length
$L=2.8\,\mu$m. As shown in Fig.\,\ref{dbl_noise}A, the BLG device
also exhibits linear $I$-$V_d$ characteristics and current
fluctuations with a 1/f frequency dependence (see inset). In
Fig.\,\ref{dbl_noise}A, one important observation is that the 1/f
noise amplitude of the BLG is found to be higher for the gate
voltage corresponding to a lower-resistance state, in sharp contrast
to the SLG device (see Fig.\,\ref{sgn_noise}A). In
Fig.\,\ref{dbl_noise}B, we plot the measured resistance and noise
amplitude $A_N$ of the BLG device as a function of $V_g$, showing a
strikingly different noise behavior from that of SLG devices. In
Fig.\,\ref{dbl_noise}B, the noise amplitude $A_N$ is minimal at the
Dirac point and increases with decreasing resistances, whereas in a
SLG device, $A_N$ is at its maximum near the Dirac point (see
Fig.\,\ref{sgn_noise}B). Fig.\,\ref{dbl_noise}C shows the noise
amplitude $A_N$ versus the resistance of the BLG device, displaying
this distinct inverse correlation between $A_N$ and $R$. It is
important to note that for most bulk semiconductor materials, if not
all, the noise amplitude $A_N$ is found to rise with increasing
resistance under field-effect modulation. Qualitatively, this is
attributed to the relatively weak dependence of the fluctuation
mechanism responsible for the noise compared to the carrier density
modulation due to the field-effect gating. To the best of our
knowledge, the noise dependence of BLG devices shown here (see
Fig.\,\ref{dbl_noise}C) is distinct from that of other known
electrical systems including carbon nanotubes, and this unique
phenomenon must be associated with the unusual band structure of the
bilayer graphene. In addition to the different gate dependence, the
1/f noise amplitude $A_N$ of the BLG device, ranging between
1$-$2$\times$10$^{-7}$ (see Fig.\,\ref{dbl_noise}B), is
significantly smaller than that of the SLG device ($\ge 10^{-6}$ in
Fig.\,\ref{sgn_noise}B), even after taking into account the factor
of $\sim 2.7$ due to the length difference. This lower $A_N$ is
consistent with the smoother $I$-$V$ curves of the BLG device shown
in Fig.\,\ref{device_setup}C.

\begin{figure} \center
\includegraphics[width=4in]{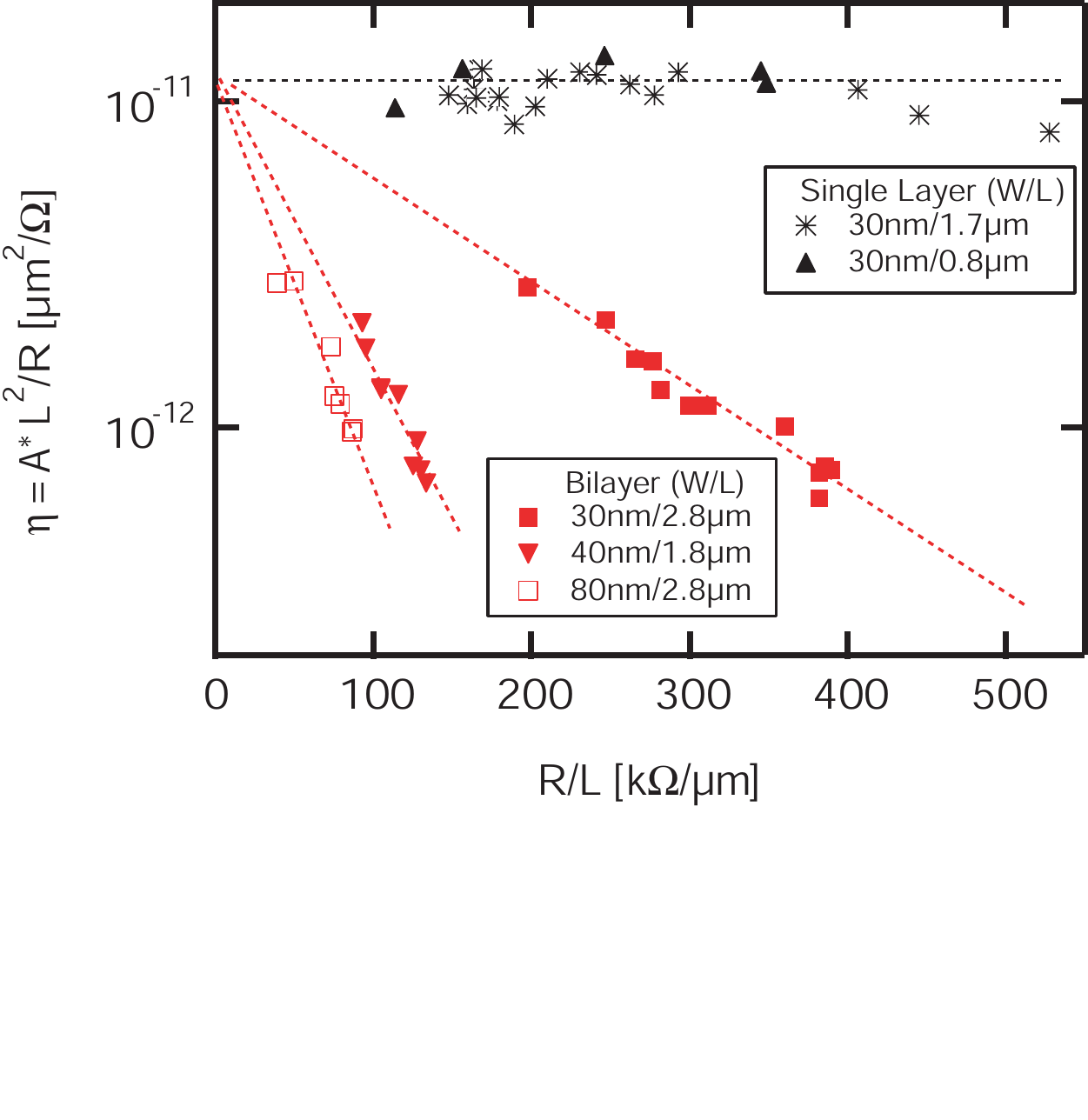}
\caption{\label{normal_compare}}
\end{figure}

To further analyze the 1/f noise characteristics observed in SLG and
BLG devices, we have fabricated and characterized both types of
devices with different channel dimensions, all showing the same
trends discussed above (see Fig.\,\ref{normal_compare}). We note
that while the unusual noise behavior in bilayer graphene does not
necessarily exclude the  validity of Hooge's relation for this
system, the
 noise parameter $\alpha_H$ cannot be unambiguously obtained in this
context using Eq.\,(\ref{eq:2}). Therefore, to compare the noise
level of graphene devices with different channel lengths and widths,
we introduce a parameter $\eta\equiv A_N L^2/R$ to describe the
noise magnitude. From Eq.\,(\ref{eq:2}), it is straightforward to
show that $\eta$ provides a convenient metric to measure the
material-specific 1/f noise level independent of device geometry,
and in SLG devices, $\eta=e\mu \alpha_H$ is directly related to the
Hooge's parameter. Fig.\,\ref{normal_compare} plots the noise factor
$\eta$ as a function of unit-length resistance ($R/L$) for two SLG
and three BLG devices with various channel dimensions. We find that
the two SLG devices yield a constant $\eta$ with comparable
magnitudes, within a factor of 2, as a function of resistance,
suggesting the same Hooge's parameter $\sim 10^{-3}$ for both
samples. On the other hand, the noise factor $\eta$ of all the BLG
devices shown in Fig.\,\ref{normal_compare} exhibits a strong
dependence on resistance, where $\eta$ increases with decreasing
$R$.  The minimal $\eta$ of the BLG devices, which is more than one
order of magnitude lower than those of SLG devices, occurs at the
Dirac point corresponding to $R_{\rm max}$.  In
Fig.\,\ref{normal_compare}, it is interesting to note that the noise
factor $\eta$ of all BLG devices, despite their different
dimensions, seems to asymptotically reach the same value as
$R\rightarrow 0$, and can be phenomenologically fitted by
\begin{equation}
\eta \sim \eta_0 \exp(-g \frac{R}{R_{\rm max}}), \label{eq:3}
\end{equation}
where $\eta_0$ is the constant noise factor in SLG and $g$ is a
constant.

Qualitatively, the asymptotic behavior of Eq.\,(\ref{eq:3}) suggests
that, in both monolayer and bilayer systems, the 1/f noise is due to
the same fluctuation mechanism, i.e.\ interaction between oxide
surface and the adjacent graphene layer. In order to evaluate
possible causes for the distinct gate dependence and noise amplitude
observed in SLG and BLG nanodevices, we first consider the carrier
concentration and charge distribution in graphene bilayers under the
influence of a gate field. The total carrier concentration $n$
induced by the applied gate voltage in an undoped graphene device is
determined by $n=C_g\cdot V_g/e$, where $C_g$ is the gate
capacitance consisting of the quantum capacitance $C_Q$ and the
electrostatic capacitance $C_e$ in series, i.e.\
$C_g^{-1}=C_Q^{-1}+C_e^{-1}$. For a typical carrier density of
$n\sim10^{12}$\,cm$^{-2}$, $C_Q$ is on the order of
$10^{-5}$\,F/cm$^{-2}$ for both graphene
monolayers\,\cite{guo_NANOLETT2007} and bilayers, i.e.\ much larger
than the electrostatic capacitance $C_e\simeq 10^{-8}$\,F/cm$^{-2}$
for 300\,nm thick SiO$_2$ dielectrics used here. Thus, for both
types of devices studied here, the gate capacitance $C_g$ is
dominated by the electrostatic capacitance $C_e$, and the total
carrier concentration is given by $n=\gamma\cdot V_g$
($\gamma\approx 7.2\times
10^{10}$\,cm$^{-2}$/V)\,\cite{novoselov_nature2005,Zhang_Kim_Nature2005}
as mentioned earlier. In graphene bilayers, we denote by $n_1$ and
$n_2$ the carrier density in the bottom (closer to the oxide ) and
the top (facing vacuum) layers, respectively, and $n=n_1+n_2$. In
the absence of screening, the two layers possess equal charge
density $n_1=n_2=n/2$, whereas in the case of perfect screening by
the bottom layer, we have $n_1=n$ and $n_2\simeq0$. In reality,
since the screening length of graphite in the c-axis is $\sim
5$\AA,\cite{graphite_screening_PRL1978} comparable to the interlayer
distance $\sim 3.4$\AA, the actual charge distribution is between
the two extremes,\,\cite{Biased_bilayer_castroPRL2007} and we have
$n_1=n/2+\Delta/2$ and $n_2=n/2-\Delta/2$, where $\Delta=n_1-n_2$ is
the charge imbalance due to the screening.  It is important to point
out that while it may seem intuitive to view the bilayer system as a
parallel-plate capacitor and attribute the lower noise level $\eta$
in BLG devices to the charge distribution $n_1 < n$, a closer
examination of this simple electrostatic screening model reveals
significant discrepancy, both qualitative and quantitative, between
the predicted noise behavior and the experimental results shown in
Fig.\,\ref{normal_compare}, as explained here. Assuming a noiseless
transport behavior for carriers residing on the {\it top} graphene
layer and similar fluctuations experienced by carriers in the {\it
bottom} layer as those in SLG, the noise factor of the BLG can be
written as $\eta=\frac{n_1}{n}\eta_0$ in the context of Hooge's
noise relation. It can be readily seen that this simple model yields
a lowest possible noise factor of only $\sim \eta_0/2$ for graphene
bilayers in the absence of screening ($n_1=n_2$), and this value is
orders of magnitude larger than our measured results (see
Fig.\,\ref{normal_compare}). Furthermore, McCann has calculated the
carrier density inhomogeneity in graphene bilayer using a
self-consistent tight-binding model and the Hartree
approach,\,\cite{McCann_PRBRapid2006} and showed that as $V_g$
varies, the charge density $n_1$ (or $n_2$) exhibits a quasi-linear
dependence on the total carrier density $n$ (up to $n\le
10^{13}$\,cm$^{-2}$). Based on the simple screening model, this
linear dependence of $n_1$ and $n_2$ on $n$ would lead to a noise
factor $\eta$ that is roughly constant as a function of $V_g$, which
is also inconsistent with the trends observed in
Fig.\,\ref{normal_compare}.

In light of the inadequacy of the simple capacitor model that only
considers the effect of the charge distribution under a given
external field, we conclude that the anomalous noise behavior
observed in BLG devices must be associated with the coupling between
the two layers that gives rise to a unique field-dependent band
structure in graphene bilayers. Both
theoretical\,\cite{McCann_PRBRapid2006} and experimental studies
have shown that an energy gap $\Delta_g$ between the conduction and
valence band can be induced in graphene bilayers by
doping\,\cite{Ohta_science2006} or applying a gate
field.\,\,\cite{Biased_bilayer_castroPRL2007,gate_induced_bilayer_nature_mat2007}
For undoped devices, the band gap $\Delta_g$ is found to linearly
increase  with the charge concentration
$n$,\,\cite{McCann_PRBRapid2006} yielding $\Delta_g\sim$10\,meV for
$n\sim 10^{12}$\,cm$^{-2}$. Therefore, for our BLG devices, the band
gap $\Delta_g$ is proportional to the gate voltage $V_g$,
$\Delta_g/V_g\simeq 13.9$\,(meV/V), and at $V_g=\pm 20$\,V, the
field-induced band gap is estimated to be $\Delta_g\simeq$ 28\,meV.
We note that in the nano-ribbon devices, a band gap
$\Delta_Q=2\times\hbar^2/(2m_eW^2)$ can also be produced by quantum
confinement effects, where $m_e$ is the electron effective mass.  To
evaluate $\Delta_Q$ in BLG nanoribbon  devices , we take
$m_e=0.05\,m_0$ ($m_0$ is the free electron mass), which is the
lower limit of measured cyclotron mass of electrons in graphene
bilayers,\,\cite{McCann_PRBRapid2006,Biased_bilayer_castroPRL2007}
and obtain $\Delta_Q\simeq 0.7$\,meV for $W=30$\,nm, indicating that
this size-induced band gap is negligible compared to the
gate-induced $\Delta_g$ in our BLG devices for most $V_g$. Similarly
in SLG nanodevices, there may also exist a size-dependent band gap
due to quantum confinement, which can be as large as 20\,meV for a
30-nm-wide ribbon.\,\cite{Han_PRL2007} We note that this band gap in
SLG devices, while independent of the gate field, is found to be
highly sensitive to the exact channel edge configurations (i.e.\
zigzag vs. armchair).\,\cite{PhysRevB.54.17954} Nevertheless, both
previous noise studies on SLG
nanoribbons\,\cite{zhihong_graphene2007} and our results here
(Fig.\,\ref{normal_compare}) reveal that the noise amplitude
$\alpha_H$ in SLG devices is not affected by this size-induced band
gap as a function of device $W$ down to 20\,nm, and therefore, the
SLG nanoribbon can be treated as a regular semiconductor in this
context.

On the basis of the above findings, a qualitative understanding of
the data shown in Fig.\,\ref{normal_compare} emerges. First, since
1/f noise mainly originates from the fluctuating trap charges in the
oxide that modulate the carrier mobility in the channel, the noise
amplitude is highly dependent on the effectiveness of the impurity
charge screening in the channel. In a graphene monolayer, the
in-plane screening of an external impurity charge is found to
exhibit a slow algebraic decay with a characteristic length of
3.8\AA\ due to both the Dirac-like dispersion relation and its
two-dimensional nature.\,\cite{Divincenzo_PRB1984} In graphene
bilayers, we expect this screening to be much stronger due to a more
{\it bulk}-like character, as implied by a comparable screening
length in the c-axis, as well as the  parabolic band-structure
distinct from that of graphene monolayers. More importantly, in
narrow gap semiconductors, the screening strength diminishes with
increasing bandgap, and therefore, the mobility fluctuation due to
the same trap charge perturbation is expected to become larger as
$\Delta_g$ increases, consistent with the trend observed in
Fig.\,\ref{normal_compare} that exhibits a minimal noise factor
$\eta$ at the Dirac point. This gate-dependent (or gap-dependent)
screening property in graphene bilayers is also reflected in the ab
initio calculation by Min et al.\,\cite{MacDonald_PRB2007} where it
is found that as $\Delta_g$ approaches 0, the ratio between the
induced channel potential variation and the external perturbing
potentials also vanishes. We believe that the low noise level
observed in the BLG, especially near the Dirac point, is related to
this dynamic charge redistribution that provides effective screening
to the nearby trap charges in the oxide.  A detailed theoretical
model for BLG that considers the self-consistent charge distribution
and the impact of screening on Columb scatters in the oxide is,
however, required in order to provide quantitative insight into the
suppressed 1/f noise in the bilayer graphene system.

In summary, we have performed 1/f noise measurements on both
single-layer and bilayer graphene nano-devices, revealing distinct
noise characteristics for the two systems. The 1/f noise in graphene
monolayers is found to follow Hooge's empirical relation, yielding a
noise level per carrier comparable to carbon nanotube devices and
bulk semiconductors despite its 2D nature. Unexpectedly, the 1/f
noise level in graphene bilayers is strongly suppressed compared to
their monolayer counterparts, and also exhibits an unusual
dependence on the carrier density in contrast to other known
materials. This unique 1/f noise behavior observed in graphene
bilayers is likely associated with a band structure that depends on
the charge distribution among the two layers, resulting in an
effective screening of scattering due to external impurity charges.
Nevertheless, both graphene monolayer and bilayer devices exhibit
1/f noise levels independent of carrier type. While further detailed
quantitative analysis and studies are required to understand these
phenomena, the findings here provide exciting opportunities for
graphene bilayers in low-noise applications.

The authors are indebted to Z.\ Chen for assistance with graphene
device fabrication and V.\ Perebeinos for useful discussions. We
also thank B.\ Ek for expert technical assistance.

\section{methods} Ultra-thin graphite films containing few-layer
graphene were obtained by mechanical exfoliation and transferred to
a highly doped Si substrate covered with 300-nm thick SiO$_2$. The
absolute number of graphene layers is determined by atomic force
microscopy (AFM) height measurements to identify single-layer and
bilayer graphene domains. To Fabricate graphene nano-devices, Pd
contacts are first deposited as the source and drain electrodes by
e-beam lithography and lift-off. Narrow ribbons of graphene between
the metal contacts are formed by oxygen plasma RIE etching using a
patterned HSQ (hydrogen silsesquioxane) layer as the protective
mask. Finally, the HSQ layer is removed in buffered HF solution. All
the electrical measurements in this report were performed under high
vacuum ($\sim 10^{-7}$\,Torr) at room temperature.


\newpage
\section*{Figure Caption}
\begin{figure} \center
\includegraphics[width=4in]{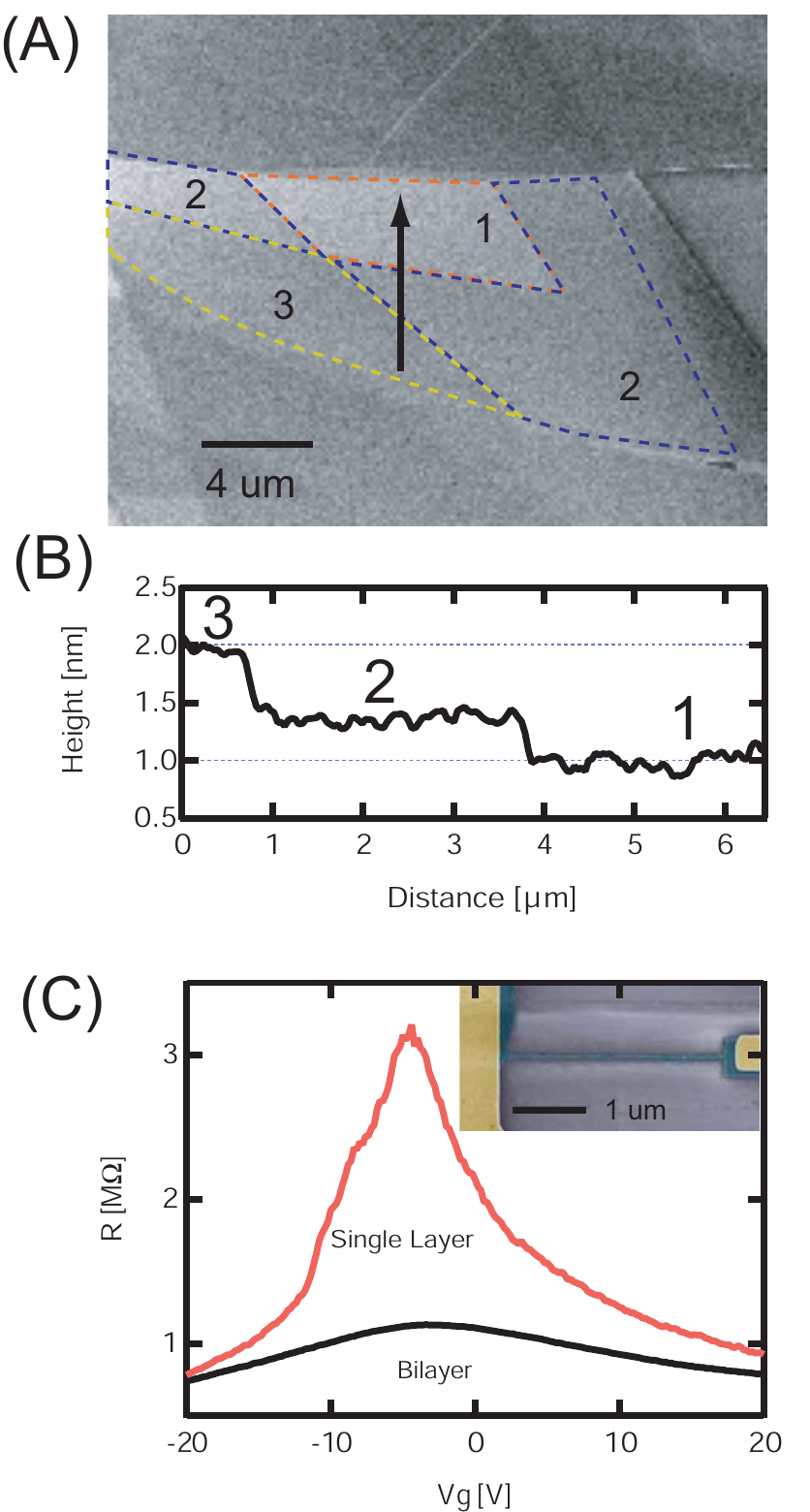}
\caption{\label{device_setup}}
\end{figure}

Fig.\ 1: \textbf{Identification of single-layer and bi-layer
graphene regions and device fabrication.} \textbf{A} SEM image of a
graphite flake deposited on the SiO$_2$/Si surface. The contrast of
the image brightness reflects segments corresponding to different
layer thicknesses, as indicated by the number of layers in the
figure. Single-layer and bilayer nanodevices in this study are
fabricated using the top and bottom graphene demains labeled by 1
and 2, respectively. \textbf{B} Trace of the height profile measured
by AFM along the arrow shown in Fig.\,\ref{device_setup}A, yielding
a height difference $\leq 5$\,\AA\, for each layer. \textbf{C}
Resistance of one single-layer and one bilayer graphene nanoribbon
devices measured as a function of gate voltage. The two devices
possess identical channel layout ($W=30$\,nm and $L=2.8\,\mu$m) as
shown in the inset.\\

Fig.\ 2: \textbf{Electrical transport and noise characterization of
a single-layer graphene nanoribbon device.} \textbf{A} Device
current as a function of drain voltage of a SLG nanoribbon device at
$Vg=$0 and -20\,V, showing the excellent linear dependence. The
inset shows the noise power spectra density $S_I$, normalized by
$I^2$, as a function of frequency (f) for two gate voltages. The
noise power spectrum at both gate voltages follows the f$^{-1}$
dependence, as indicated by the solid lines, and is called the 1/f
noise. Note that at $V_g$=-20\,V the device exhibits a smaller
resistance as well as a lower noise level $S_I/I^2$ (dashed curve in
the inset). \textbf{B} The resistance and the noise amplitude $A_N$,
defined in Eq.\,(\ref{eq:noise}), of the SLG nanoribbon device
measured as a function of gate voltage. The dashed curve is a guide
to the eye, illustrating the correlation between $A_N$ and $R$.
\textbf{C} The noise amplitude $A_N$ and the frequency exponent
$\beta$ of the SLG device plotted as a function of resistance.  The
frequency exponent $\beta \simeq -1$ confirms the 1/f noise behavior
for the gate voltage range studied, and thus, ensures the validity
of the obtained $A_N$ . The solid line is a linear fit of $A_N$
versus $R$,
yielding good agreement with Hooge's empirical relation.\\

\begin{figure} \center
\includegraphics[width=4in]{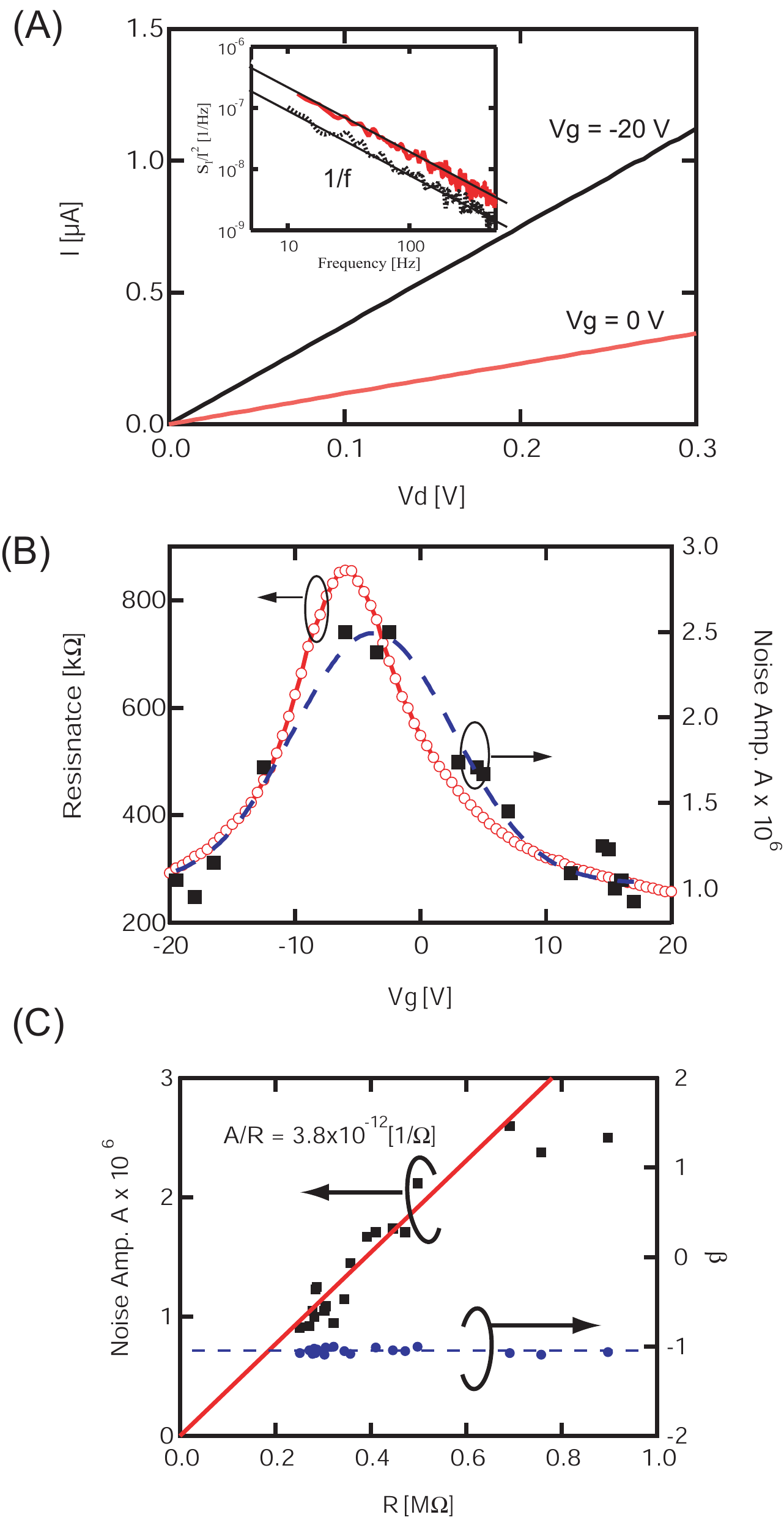}
\caption{\label{sgn_noise}}
\end{figure}

Fig.\ 3: \textbf{Electrical transport and noise characterization of
a bilayer graphene nanoribbon device.} \textbf{A} Device current as
a function of drain voltage of a BLG nanoribbon device at $Vg=$0 and
-20\,V, showing the linear dependence. The inset shows noise power
spectra $S_I/I^2$ as a function of frequency for these two gate
voltages. At $V_g$=-20\,V, the BLG device possesses a smaller
resistance, but exhibits a higher noise level (dashed curve in the
inset), which is in contrast to the case of a SLG device shown in
Fig.\,\ref{sgn_noise}A. \textbf{B} The resistance and the noise
amplitude $A_N$ of the BLG device measured as a function of gate
voltage. The dashed curve is a guide to the eye, illustrating the
inverse relation between $A_N$ and $R$. \textbf{C} The noise
amplitude $A_N$ and the frequency exponent $\beta$ of the SLG device
as a function of resistance.  While the frequency exponent still
yields the expected $\beta \simeq -1$, the noise amplitude $A_N$
increases with decreasing $R$, in drastic contrast to the SLG device
shown in Fig.\,\ref{sgn_noise}C. The dashed line on $A_N$ is a guide
to the eye to highlight this difference.\\

Fig.\ 4: \textbf{Comparison of the noise characteristics of SLG and
BLG devices.} The measured noise factor $\eta$, defined as $A_N
L^2/R$, as a function of unit-length resistance for two SLG and
three BLG devices with different channel dimensions. Both SLG
devices shown here exhibit a nearly constant noise factor $\eta$ as
the resistance varies, in consistent with Hooge's relation, and
yield $\alpha_H\sim 10^{-3}$. In contrast, all three BLG devices
possess a much lower noise factor $\eta$ than that of SLG devices.
Moreover, these BLG devices possess a noise factor $\eta$ that rises
with decreasing resistance, which can be phenomenologically
described by Eq.\,(\ref{eq:3}).\\

\end{document}